\documentclass[twocolumn,aps,footinbib,pra]{revtex4}
\usepackage{amssymb}
\usepackage{amssymb}
\usepackage{graphicx}
\usepackage{amsmath}
\usepackage{times}
\usepackage{color}

\begin{document}

\title{ Quantum limited measurements of atomic scattering properties }
\author{ A. M. Rey$^{1}$, L. Jiang$^{2}$ and M. D.
Lukin$^{1,2}$}  \affiliation{$^{1}$ Institute for Theoretical
Atomic, Molecular and Optical Physics, Cambridge, MA, 02138.}
\affiliation{$^{2}$ Physics Department, Harvard University,
Cambridge, Massachusetts 02138, USA}
\date{\today }

\begin{abstract}
We propose a method to perform precision measurements of the
interaction parameters in  systems  of  $N$  ultra-cold  spin $1/2$
atoms. The spectroscopy is realized by first creating a coherent
spin superposition of the two relevant internal states of each atom
and then letting  the atoms evolve under a squeezing Hamiltonian.
The non-linear nature of the Hamiltonian decreases the fundamental
limit imposed by the Heisenberg uncertainty principle to $N^{-2}$, a
factor of $N$ smaller than the fundamental limit achievable with
non-interacting atoms. We study the effect of decoherence and show
that even with decoherence, entangled states can outperform the
signal to noise limit of  non-entangled states. We present two
possible experimental implementations of the method using
Bose-Einstein spinor condensates and fermionic atoms loaded in
optical lattices and discuss their advantages and disadvantages.
\end{abstract}


\maketitle

\section{Introduction}

The preparation of many-particle entangled states is becoming a
fundamental task in modern quantum physics. Entanglement  lies at
the heart of quantum communication and quantum information and is
also  a fundamental resource in
 precision spectroscopy.  With recent
advances in the manipulation of   trapped ions and neutral atoms,
there has been significant progress in the preparation  of many-body
entanglement states  in  atomic gases
\cite{Polzik,Leibfried,Leibfried2,Mandel}. Such systems therefore
offer rich opportunities to  investigate new   physics with both
practical and fundamental applications.

Here we propose an interferometric method that uses the entanglement
that emerges during the many-body dynamics between interacting
spinor atoms to perform precision measurements of their interatomic
interaction strength, usually parameterized by the $s$-wave
scattering length. Our method is a many-body generalization of the
technique reported in Ref. \cite{Widera1}, where  information of the
scattering properties was obtained  by  using the entanglement
dynamics in pairs of atoms trapped in the ground state of a
potential well.

A precise determination of the scattering properties   has broad and
important applications beyond atomic physics. For example,
Ref.~\cite{Flambaum} shows that such type of measurements might be
useful for fundamental physics,  as by monitoring the scattering
length on the $10^{-2} $ level one could detect variation in the
electron to proton mass ratio  on the level of $10^{-11}- 10^{-14}$.

Standard Ramsey spectroscopy starts with a system of $N$
non-interacting spin $1/2$ particles initially prepared in the same
internal state. Subsequently, a Ramsey pulse prepares it  in a
coherent superposition of the two internal states and  the system is
let to evolve freely for a  time during which a relative phase
accumulates, $\phi$, due to the energy splitting between the  two
states. The latter is then  decoded by using a second Ramsey pulse
which maps the phase onto a population difference, thus allowing to
perform precision measurements of the atomic transition frequency.

The statistical fluctuations associated with a finite sampling
yields a lower  limit in the phase accuracy $\delta \phi=\sqrt{1/N}$
called the shot noise limit \cite{Itano}. The Heisenberg uncertainty
principle, however,  allows for phase accuracies consistent with the
basic principles of quantum mechanics, as low as  $\delta \phi=1/N$,
called the Heisenberg limit. The latter can be achieved by using
initially entangled atoms instead of a polarized sample \cite{
Bollinger}.

Here we propose a spectroscopy technique which  aims to measure the
atomic interaction parameters arising form two-body collisions by
using similar Ramsey spectroscopic ideas, but applied to interacting
atoms instead of non-interacting ones. The many-body interactions
during the free evolution build in quantum mechanical correlations
between the atoms and generates a large amount of entanglement even
from an initially uncorrelated sample \cite{Sorensen}. This   scheme
provides a resolution which is fundamentally limited to $N^{-2}$,  a
factor of $N$ smaller than the Heisenberg  limit achievable with
non-interacting atoms. Moreover, for initially uncorrelated atoms
the sensitivity scales as $N^{-1}$, implying a gain in resolution by
a factor of $\sqrt{N}$ with respect to the classical shot noise
resolution. A  more general analysis of
 beyond-Heisenberg scaling for multi-body collisions or tensor-field interactions has been
 derived in Ref. \cite{Boixo}.

Many-particle entangled states, however,  are difficult to prepare
and maintain since they are extremely fragile: in practice, noise
and decoherence rapidly collapse entangled states into classical
statistical mixtures. For example in standard Ramsey spectroscopy
decoherence destroys the potential gain provided by entanglement,
and,  when  it is taken into account  both initially uncorrelated
and maximally entangled states provide the same shot noise
resolution \cite{Huelga}. Here we show that decoherence  degrades
the sensitivity of our method to $N^{-3/2}$ for initially entangled
atoms and  to $N^{-3/4}$ for initially uncorrelated ones. This
scaling is interesting as it demonstrates that   the
 entanglement build up by the many-body interactions helps,  even with decoherence,
to keep the sensitivity achievable with initially squeezed atoms
above the maximally achievable with initially uncorrelated
particles.

We also  discuss physical implementation of our scheme using a)
 spinor Bose- Einstein condensates and b) fermionic atoms loaded in optical
lattices and  compare and contrast the advantage and disadvantages
of these two possible set-ups. The paper is organized as follows. In
Sec. II we present our interferometry method based on the many-body
collective dynamics. In Sec. III we apply it to initially
uncorrelated states and   derive the type of initially entangled
states that lead to Heisenberg-limited sensitivity.  In this section
we  also propose a way to generate such states. In Sec. IV we
include the effect of decoherence. In  Sec. V we discuss the
experimental implementation of the interferometry method using
spinor condensates and cold fermionic atoms in optical lattices and
discuss possible technical limitations. Finally we conclude in Sec.
VI.

\section{Interferometric technique}

Let us consider a collection of $N$ spin-1/2 interacting particles described
by the Hamiltonian

\begin{equation}
\hat{H}_{m}=\chi \hat{J}_{z}^{2}.
\end{equation}
 We used $\hat{J}%
_{\alpha }$ to denote the collective spin operators of the $N$
atoms: $\hat{J} _{\alpha }=\sum _i\hat{\sigma}_{i}^{\alpha }$, where
$\alpha =x,y,z$ and $\hat{\sigma}_{i}^{\alpha }$ is a   Pauli
 operator acting on the   $i^{th}$ atom. We set $\hbar
=1$.

Our interferometric technique  follows the same ideas used in
standard Ramsey spectroscopy, but replaces the free evolution of the
atoms by evolution with  $\hat{H}_{m}$ and the goal is to estimate
$\chi $ as accurately as possible. We start with the initial state
$|\Phi _{0}\rangle $ and apply a $\pi /2$ rotation
to all particles about the y axis, $|\psi (0)\rangle =e^{i\pi /2\hat{J}%
_{y}}|\Phi _{0}\rangle $. After letting the system evolve for time
$t$  under $\hat{H}_{m}$, $\left\vert \psi (t)\right\rangle =e^{-it%
\hat{H}_{m}}|\psi (0)\rangle $, a second $-\pi /2$ pulse is applied,
$\left\vert \psi_f (t)\right\rangle =e^{-i\pi /2\hat{J}_y}|\psi
(t)\rangle $. Finally the collective spin is measured $\langle
\hat{J}_{z}(t)\rangle= \langle \psi_f (t)| \hat{J}_{z}|\psi_f
(t)\rangle $ . If this  scheme is repeated during a total time $T$
the achievable sensitivity is given by
\begin{equation}
|\delta \chi |^{2}=\frac{t}{T}\frac{\langle \Delta \hat{J}_{z}(t)\rangle }{%
(\delta \langle \hat{J}_{z}(t)\rangle /\delta \chi )^{2}}=\frac{t}{T}\frac{%
\langle \psi (t)|\Delta \hat{J}_{x}|\psi (t)\rangle }{(\delta
\langle \psi (t)|\hat{J}_{x}|\psi (t)\rangle /\delta \chi )^{2}},
\label{sig}
\end{equation}
with variance $\Delta \hat{J}_{\alpha }=\hat{J}_{\alpha }^{2}-\langle \hat{J}%
_{\alpha }\rangle ^{2}$.

 In contrast to standard Ramsey spectroscopy, which is fundamentally limited by the
 Heisenberg exclusion principle to $ {(tN)}^{-1}$\cite{Huelga,
Bollinger} per realization, the fundamental limit  of this scheme is
$ 2/(tN^{2})$. This follows from an application of the time-energy
uncertainty principle: $\delta t^{2}\langle \Delta
\hat{H}_{m}\rangle \geq 1/4$ where $\langle \Delta
\hat{H}_{m}\rangle $ is the variance of the Hamiltonian and $\delta
t^{2}$ is the variance in estimating time from a measurement on the
system \footnote{Measurements of an operator $\hat{A}$ can be used
to determine time with an uncertainty $\sqrt{\langle \Delta
\hat{A}}\rangle /d\hat{\langle A\rangle } /dt$}. The time-energy
uncertainty can be reexpressed as $\delta \varphi ^{2}\langle \Delta
\hat{J}_{z}^{2}\rangle \geq 1/4$, where $\varphi =\chi t$. We can
establish an upper limit $\langle \Delta \hat{J}_{z}^{2}\rangle $
from $\langle \Delta \hat{J}_{z}^{2}\rangle =\langle
\hat{J}_{z}^{4}\rangle -\langle \hat{J}_{z}^{2}\rangle ^{2}\leq
\langle N^{4}/16\rangle $\footnote{ The  inequality follows because
the maximum eigenvalue of $\hat{J}_{z}^{4}$ is $N^{4}/16$}. This
implies $\delta \varphi \geq 2N^{-2}$ and therefore a fundamental
uncertainty in determining $\delta \chi $ after a time $t$ of: $
|\delta \chi |\geq 2/(tN^{2})$.

\section{Spectroscopy}

Let us  start by outlining the basic properties of $\hat{H}_{m}$. It
 commutes with $\hat{J}^{2}$ and $\hat{J}_{z}$, so a
good basis to describe the dynamics is the one  spanned by the
collective angular momentum eigenstates. In the $2^{N}$ dimensional
Hilbert space, there are $ N+1$ orthogonal fully symmetric states
which we denote by $|J=\frac{N}{2},M\rangle _{z}$ that satisfy:
$\hat{J}^2 |\frac{N}{2},M\rangle_z=\frac{N}{2}
(\frac{N}{2}+1)|\frac{N}{2},M,\rangle_z$, and
$\hat{J}_z|\frac{N}{2},M\rangle_z= M |\frac{N}{2},M\rangle_z$ with
$-\frac{N}{2}\leq M \leq \frac{N}{2}$. We refer to the $J=N/2$
states as the $\mathcal{P}$ manifold. The quantum numbers $J$ and
$M$ are conserved during the  dynamics, and therefore, if the
initial state $\left\vert \Phi _{0}\right\rangle $ belongs to
$\mathcal{P}$, then the subsequent evolution of the system only
takes place within $\mathcal{P}$.

\subsection{Initially uncorrelated atoms}

\begin{figure}[tbh]
\begin{center}
\leavevmode {\includegraphics[width=2.5 in]{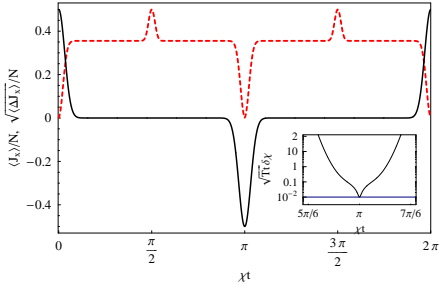}}
\end{center}
\caption{ $\langle \hat{ J}_x\rangle_\psi$/N (solid black line) and
$ \sqrt{\langle\Delta\hat{J}_x}\rangle_\psi$ /N (dashed red line) as
a function of time  for a initially polarized state with $N=100$. In
the inset we plot  the corresponding  sensitivity. The horizontal
line is at $|\delta\chi|_{opt}$ given by
Eq.(\ref{seno}).}\label{fig1}
\end{figure}

First, consider the case when  the initial state is  fully polarized
 along $z$: $ |\Phi _{0}\rangle =|\frac{N}{2},\frac{N}{2}
\rangle_z$. After the first $\pi /2$ pulse it becomes $|\psi
(0)\rangle =|\frac{N}{2},\frac{N}{2}\rangle _{x}\approx
\sum_{M=-N/2}^{N/2}\left( \frac{4}{2\pi N}\right)
^{1/4}e^{-M^{2}/N}|N/2,M \rangle _{z}$. The effect of $\hat{H}_{m}$
on the state is to imprint a phase $\chi M^{2}t$ to each $\left\vert
N/2,M\right\rangle _{z}$ component. In the limit of many atoms
$(N\gg 1)$ we may approximate the sums by integrals assuming $M/N$
is a continuous variable  and to replace the binomial coefficient by
a Gaussian distribution with the same width.  In this limit $\langle
\hat{J}_{x}\rangle_\psi $ and $\langle \hat{J}^2_{x}\rangle_\psi$
become
\begin{eqnarray}
&&\langle \hat{J}_{x}\rangle_\psi  =\frac{N}{2}
\sum_{k=0,1,2,\cdots }(-1)^{k}e^{-N/2(\chi t-k\pi )^{2}},  \label{jz} \\
&&\langle \hat{J}^2_{x}\rangle_\psi =\frac{N^{2}-N}{8}
\sum_{k=0,1,2,\cdots }e^{-N(\chi t-k\pi /2)^{2}}+\frac{N^{2}+N}{8},
\label{jz2}
\end{eqnarray} where we used the notation  $\langle\hat{A}\rangle_\psi\equiv
\langle \psi(t)|\hat{A}|\psi(t)\rangle $. The time evolution of the
signal is a series of equally spaced gaussian pulses (see Fig.
\ref{fig1}). As the system evolves, at  first the different
accumulated phases  lead to a collapse  of $\langle
\hat{J}_{x}\rangle_\psi$. For the initial coherent state  in
consideration. the collapse time, which depends on its variance, is
$\chi t_{{\rm coll}}\approx 3 N^{-1/2}$. The evolution continues and
at $\chi t_{rev}= \pi $ all the different components rephase
inducing a perfect revival of the initial state, now with opposite
polarization:$\langle \hat{J}_{x}\rangle_\psi=-N/2$. Using Eqs.
(\ref{jz}) and (\ref{jz2}) one can show that the optimal sensitivity
is reached at the points  with maximal signal intensity, $\chi
t=k\pi $. It scales as $N^{-1}$ (see Fig. \ref{fig1}):
\begin{equation}
|\delta \chi |_{opt}=\frac{1}{N\sqrt{Tt}}. \label{seno}
\end{equation}

\subsection{Initially correlated atoms}

\begin{figure}[tbh]
\begin{center}
\leavevmode {\includegraphics[width=3. in]{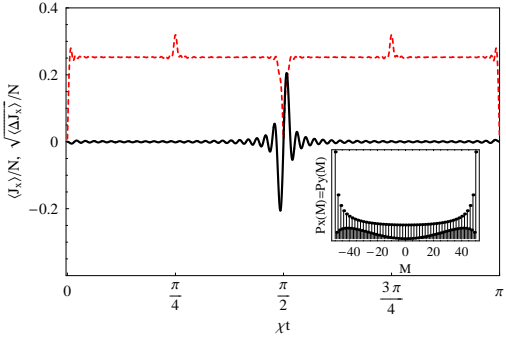}}
\end{center}
\caption{ $\langle \hat{ J}_x\rangle_\psi^a$/N (solid black line)
and $ \sqrt{\langle\Delta\hat{J}_x}\rangle_\psi^a$ /N (dashed red
line) as a function of time for an initially squeezed state given by
Eq.(6) with  $N=100$ and $a=0.5$. In the inset we plot the
projection probability of $|\Phi(0)^a\rangle$ onto the
$|N/2,M\rangle_{x,y}$ basis. }\label{fig2}
\end{figure}

Entangled atomic states (e.g. spin squeezed states) potentially
allow to significantly improve the sensitivity in precision
measurements. For example, consider  the initial
state\begin{equation}
|\Phi^a (0)\rangle =\frac{\left( |\frac{N}{2},0\rangle _{z}+a\left[ \frac{|%
\frac{N}{2},2\rangle _{z}-|\frac{N}{2},-2\rangle
_{z}}{\sqrt{2}}\right] \right) }{\sqrt{1+a^{2}}} . \label{inisquee}
\end{equation} Here $a$ is a real
number of order unity and $N$ is assumed to be even and large. The
probability distribution of $\left\vert \Phi^a (0)\right\rangle $ in
the $\{|N/2,M\rangle _{x}\}$ basis   is plotted in the inset of
Fig.\ref{fig2}. For even $M$ values,  the amplitudes are almost
constant (except for those with $|M|$ close to $N/2$), and  for odd
$M$  values they are proportional to $a^{2}$. $\left\vert \Phi^a
(0)\right\rangle $ is a highly squeezed state with reduced variance
in the $z$ direction: $\langle \Phi _{0}|\Delta \hat{J}_{z}|\Phi
_{0}\rangle =\frac{4a^{2}}{1+a^{2}}\ll \langle
\Phi _{0}|\Delta \hat{J}_{x,y}|\Phi _{0}\rangle =\frac{N(N+2)}{8}-\frac{%
2a^{2}}{(1+a^{2})}$.

\begin{figure}[tbh]
\begin{center}
\leavevmode {\includegraphics[width=3.7 in]{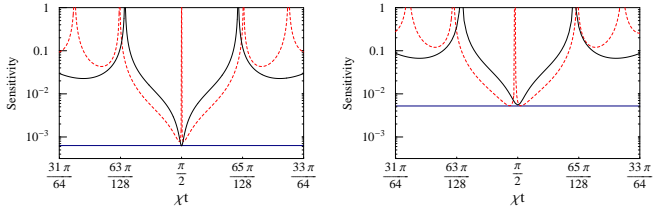}}
\end{center}
\caption{Optimal sensitivity vs time ( $\sqrt{T}$ units) for
$|\Phi(0)^a\rangle$ by measuring $\hat{J}_z$ (solid black curve) and
$\hat{J}_z^2$ (red dashed line). The left panel is without
decoherence and the  right with decoherence. The horizontal lines
are  at $|\delta\chi|_{opt}^{a}$ and $
|\delta\chi|_{opt}^{a,\Gamma}$. Here $a=0.5$ and
$N=100$.}\label{fig3}
\end{figure}

Using properties of the Wigner rotation functions, it is possible to
show that:

\begin{eqnarray}
\langle \hat{ J}_x\rangle_\psi^a&=&\frac{N a \sqrt{8}}{1+a^2}%
\sum_{k=0} \frac{J_2 [N(\chi t-\pi/2+ k \pi)]}{N (\chi t-\pi/2
+k\pi)},
\label{jzs} \\
\langle  \hat{J}_x^2\rangle_\psi^a &=& N^2A\sum_{k=0} \left(\frac{%
J_1[2N(\chi t- k \pi/2]}{(2N (\chi t- k\pi/2))}\right) +B,
\label{jz2s}
\end{eqnarray}
with $A=\frac{6 a^2}{N^2(1+a^2)}-\frac{N+2}{8N}$ and
$B=\frac{a^2}{1+a^2} +\frac{N(N+2)}{16}$. As shown in Fig.
\ref{fig2} the system starts at $t=0$ with $\langle \hat{
J}_x\rangle_\psi^a=0$, and during the time evolution $\langle \hat{
J}_x\rangle_\psi^a$ grows while oscillating. At $\chi
t\cong\frac{\pi}{2} \pm 2.29 N^{-1}$ the signal reaches a maximum
and vanishes again at $\chi t=\frac{\pi}{2}$. Then it  evolves with
opposite polarization until it returns to the initial state at $\chi
t=\pi$.  Using Eqs. (\ref{jzs}) and (\ref{jz2s}), the optimal
sensitivity  can be shown to be reached at $\chi t=k \pi/2$, with
$k$ an odd integer (see Fig.\ref{fig3}):
\begin{eqnarray}
|\delta\chi|_{opt}^{a}&=&\sqrt{\frac{32 (1+a^2)}{ T t N^4}}.
\label{signds}
\end{eqnarray} These states lead to $1/N^2$ sensitivity.
The best accuracy is obtained for $a \to 0$, in which case
$|\delta\chi|_{opt}^{a}=\sqrt{\frac{32}{ T t N^4}}$. Note, however,
that   estimating  $\chi$ by measuring $\hat{J}_z$  might be
experimentally impractical due to the fact that not only the signal
vanishes at $\chi t=k \pi/2$ (regardless of the value of $a$) but
also $|\delta\chi|$ is very sharply peaked at  $\chi t=k \pi/2$ and
hence very sensitive to small variations of $\chi t$. To overcome
this limitation we propose, following the ideas exposed in Ref.
\cite{Kasevich}, to measure $\hat{J _z^2}$ instead of $\hat{J}_z$.
In contrast to $\hat{J}_z$, the average square signal does not
vanishes for finite $a$ (it goes like $\frac{4 a^2}{1+a^2}$) and, as
shown in Fig. \ref{fig3}, measuring $\chi$ by means  of $\hat{J
_z^2}$ provides the same $N^{-2}$  accuracy with the advantage of  a
broader profile around the optimal value  which becomes slightly
shifted from $ \chi t=k \pi/2$. We calculated $|\delta\chi|$
numerically according to Eq. (\ref{sig}) but
 replacing
$\hat{J}_z$ by $\hat{J}_z^2$. This idea, however, has the drawback
that  $|\delta\chi|_{\hat{J}_z\ \to \hat{J}_z^2}$ diverges exactly
at $ \chi t=k \pi/2$. Nevertheless as shown in Fig. 3 this is a
extremely narrow divergency and  almost a one point behavior.

Let us now  discuss how to generate  states of the form given by
Eq.(\ref{inisquee}). These states  are  highly squeezed, and  a
robust method to generate them can have also applications in
Heisenberg limited phase detection \cite{Kasevich}.  The idea is to
 adiabatically convert $| \frac{N}{2},\frac{N}{2}\rangle_x $ into  $|
\frac{N}{2},0\rangle_z $  by  gradually increasing $\lambda _{t}$
from $0$ to $1$ in the following time dependent
Hamiltonian\cite{Unanyan}:
\begin{equation}
\hat{H}_{adi}\left( \lambda _{t}\right) = -\zeta \hat{J}^{2}+\left(
1-\lambda _{t}\right) \omega _{x}\hat{J}_{x}+\lambda _{t}\chi
\hat{J}_{z}^{2}. \label{Hadiabatic}
\end{equation}The first term proportional to $\zeta$ is required to keep the
system within $\mathcal{P}$. As shown in Fig. 4, there is always a
finite gap between the (non-degenerate) ground state \footnote{ The
non-degeneracy requires the particle number to be even. In the case
of odd particle numbers, the degeneracy of the ground state can be
lifted by introducing $\lambda _{t}\chi \hat{J}_{z}$.} and the
lowest excited state, so the   adiabatic passage  can be satisfied.
The experimental implementation  Eq.(\ref{Hadiabatic})  will be
discussed in Sec.VB

\begin{figure}[tbh]
\begin{center}
\leavevmode {\includegraphics[width=2.5 in]{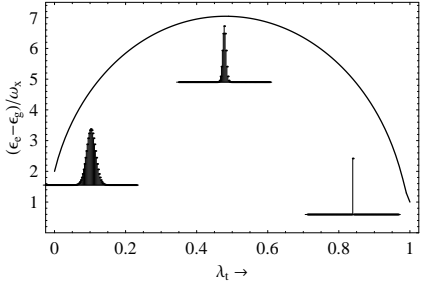}}
\end{center}
\caption{Energy gap
between the ground and the first excited states of $\hat{H}_{adi}$ vs $%
\protect\lambda _{t}$. The insets show the projection probability of the ground state into the  $J_{z}$ eigenvalues
 for $\lambda
_{t}=0,0.5$ and $1$, respectively. $\chi=1/2 \omega_x$}
\end{figure}

Once $| \frac{N}{2},0\rangle_z $ is obtained,  the $|
\frac{N}{2},\pm 2\rangle_z $ components can be generated by evolving
$| \frac{N}{2},0\rangle_z $  with the Hamiltonian $ H_{2ct}=i\xi
\left( J_{+}^{2}-J_{-}^{2}\right)$. As proposed in Ref.
\cite{Andre02}, $ H_{2ct}$ can be implemented  by using two laser
beams with frequency difference twice  the hyperfine level splitting
which coherently flips pairs of aligned spins \footnote{ For large
$a$ comparable to unity, excitations to states with $M=\pm 4,\pm
6,\cdots $, should also be considered.}.

\section{effects of decoherence}

 In realistic
experiments decoherence effects are inevitably present. The main
type of decoherence is dephasing due to processes that cause random
changes in the relative phase of the quantum states, while
preserving the total atomic population in the atomic levels. Among
the mechanisms that result in dephasing effects one can consider
collisions, stray fields, and laser instabilities. We model the
phase decoherence by adding a term to the Hamiltonian given by
$\hat{H}_{env}=\frac{1}{2}\sum_{i}h_{i}(t)\hat{\sigma}_{i}^{z}$
where $h_{i}(t)$ are independent random gaussian variables with zero
mean
and white noise correlation functions $\overline{h_{i}(t)h_{j}(\tau )}%
=2\Gamma \delta _{ij}\delta (t-\tau )$ that represent couplings with
the environment. Here the bar denotes averaging over the different
outcomes of the random variables. This model is equivalent to
considering the time evolution of the reduced density operator for a
single ion $\hat{\rho}$ as given  by the master equation
$\dot{\hat{\rho}}=i[\hat{\rho},\hat{H}_{m}]-\Gamma /2(\sigma
_{z}\hat{\rho}\sigma _{z}-\hat{\rho})$.

To include the effect of decoherence it is simpler to go to the
uncoupled basis which  diagonalizes both $\hat{H}_{env}$ and
$\hat{H}_{m}$. Each state in this basis can be label as
$\{|n^{(k)}\rangle =|s_{1}^{k},s_{2}^{k},\dots ,s_{N}^{k}\rangle
\}$, where $s_{i}^{k}=\pm 1$  and $k=1,\dots 2^{N}$. Using the
uncorrelated nature of the $h_i$ variables one can show by
straightforward calculations that \cite{Huelga}

\begin{eqnarray}
&&\langle\hat{ J}_x\rangle_{\psi,\Gamma}=e^{-\Gamma t} \langle
\hat{ J}_x\rangle_{\psi,0},  \label{dec1} \\
&&\langle  \hat{J}_x^2\rangle_{\psi,\Gamma}=e^{-2 \Gamma t} \langle
\hat{J}_x^2\rangle_{\psi,0}+ \frac{ N}{4} \left(1-e^{-2 \Gamma
t}\right),\label{dec2}\\ &&\langle \hat{J}_x^4\rangle_{\psi,\Gamma}
=e^{-4 \Gamma t} \langle \hat{J}_x^4\rangle_{\psi,0}+ g(t)
  \left(\langle
\hat{J}_x^2\rangle_{\psi,0} -\frac{N}{4}\right)+  f(t),\notag
\end{eqnarray} with $g(t)=\frac{3N-4}{2} \left(e^{-2\Gamma t}-e^{-4\Gamma
t}\right)$ and$f(t)=\frac{3N^2-2N}{16}\left(1-e^{-4\Gamma
t}\right)$. Note that these equations are valid for an arbitrary
initial state.  The noise  induces an exponential decay of  the
average signal and  its higher moments. However,  unlike the average
signal the higher moments are also affected by the presence of
diffusion terms.

If we assume an initially coherent state with $\langle
\hat{J}_x\rangle_{\psi,0}$ and
$\langle\hat{J}_x^2\rangle_{\psi,0}$  given by Eqs. (\ref{jz}) and (\ref{jz2}),  one can calculate
  from Eqs. (\ref{dec1}) and (\ref{dec2})the  sensitivity in the presence of
decoherence. In contrast to the ideal dynamics where the longer the
interrogation time $t$ the better is the phase accuracy, when
decoherence is accounted for, one has to optimize with respect to
both $t$ and $\chi t$, in order  to obtain the best sensitivity.
Provided that $T>1.6/(2 \Gamma)$, standard minimization yields:


\begin{equation}
|\delta\chi|_{opt}^{\Gamma}=\sqrt{\frac{2\sqrt{2}\Gamma}{ T N^{3/2} }\frac{%
(e^{\tau_{opt}}-1)^{1/2}}{\tau_{opt}}}=\sqrt{\frac{3.5\Gamma}{ T
N^{3/2} }}, \label{soptd}
\end{equation}at $\chi t \approx k\pi \pm \sqrt[4]{\frac{2}{N^3} (e^{\tau_{opt}}-1)}$ and
$\tau_{opt}\equiv 2 \Gamma t_{opt}  \approx 1.6$.

Note that even with initially uncorrelated atoms decoherence reduces
the sensitivity   by a factor of $N^{1/4}$. This has to be
contrasted with standard Ramsey spectroscopy  done with
non-interacting atoms, where the phase resolution with initially
uncorrelated atoms  has the same classical shot noise resolution so
is not too much affected by decoherence. The reason of the different
behavior is the fact that  $\hat{H}_m$   builds up particle
correlations during the evolution and transforms uncorrelated states
into highly squeezed ones which are vulnerable to decoherence. More
specifically, as shown in Ref. \cite{Molmer}, $\hat{H}_m$ ideally
transforms initially coherent states  into maximally correlated
$N$-particle GHZ states \cite{Greenberger} after a period of
evolution of  $\chi t =\pi/2$. These states are very fragile and
decohere $N$ times faster than uncorrelated particles.

Let us now consider the case when the initial state is
$|\Phi^a(0)\rangle$. Using  Eqs. (\ref{jzs}) and (\ref{jz2s}) in
Eqs. (\ref{dec1}) and (\ref{dec2}), one gets an optimal sensitivity,
provided $T>\tau_{opt}/(2\Gamma)$, given by:
\begin{eqnarray}
|\delta\chi|_{opt}^{a,\Gamma}&=&\sqrt{\frac{4 \Gamma (1+a^2)^2 e^{-\tau_{opt}}}{T a^2 N^3%
}},  \label{signdecma}
\end{eqnarray} at $(\chi t) =k \pi/2$, and $%
\tau_{opt}=\sqrt{\frac{2 a^2}{(1+a^2) N}}$. Eq. (\ref{signdecma}) is
very interesting as  it demonstrates that  by using many-body
interactions, entangled states can outperform the signal to noise
limit of classically unentangled states even in the presence of
decoherence. The sensitivity of the squeezed states is a factor $
N^{3/4}$ larger than the maximal achievable with uncorrelated atoms,
see Eq.(\ref{soptd}).  In Fig. 3 we  plot the optimal sensitivity in
the presence of decoherence both  by measuring $\hat{J}_z$ and
$\hat{J}_z^2$.  For practical reasons, as the ones discussed above,
 measuring $\hat{J}_z^2$ instead of $\hat{J}_z$
might be experimentally  convenient also  with decoherence.
Moreover, note that while states with $a=0$ provide the best
accuracy in the ideal dynamics, states with $a=1$ are optimal with
decoherence.

\section{Experimental implementations}

\subsection{Spinor Condensates}

$\chi \hat{J}_{z}^{2}$  describes the spin dynamics  of a $F=1$ Bose
Einstein condensate, if one  associates the internal $M_{F}=\pm 1$
hyperfine states of the atoms   as the two states of an effective
spin-1/2 particle (the $M_{F}=0$ state can be decoupled by
introducing a detuning much larger than the chemical potential)
\cite{Sorensen}. In this system the coupling strength $\chi $ is
proportional to $4 \pi /(m V) (a_{\uparrow \uparrow}+a_{\downarrow
\downarrow} -a_{\downarrow \uparrow})$ with $a_{\sigma \sigma'}$ the
s wave scattering length between particles of type $\sigma$ and
$\sigma'$, $m$ the atom mass and $V$ the volume.

 Due to the  dependence  of
$\chi$ on the  elastic  scattering length difference, $\delta a
\equiv a_{\uparrow \uparrow}+a_{\downarrow \downarrow}
-a_{\downarrow \uparrow}$, using an  inter-state Feshbach resonance
\cite{Feshbach}, such as the one predicted between the $F=2$ and
$F=1$ states in $^{87}{\rm Rb}$ atoms, one could widely tune $\chi$
by varying an external magnetic field. Close to the Feshbach
resonance there  is a singularity in the scattering length,
$a_{\downarrow \uparrow}(B)=a_{\downarrow \uparrow}^{bg} (1-\Delta
B/(B-B_o))$ with $ \Delta B $ being the resonance width, $B_o$ the
zero energy resonance and $a_{\downarrow \uparrow}^{bg}$ the
background scattering length. This singularity is due to the near
degeneracy of the collision energy of the atoms with the binding
energy of an extremely loose diatomic molecular state.

One can use our spectroscopy technique to measure $\delta a$ in such
system. However, in order for the method to work, it is important to
reduce two and three body particle losses across the resonance, as
they can considerably affect the sensitivity. We estimate for
example the effect of two-body losses by solving the following
master equation:

\begin{equation}
\dot{\hat{\rho}}=i[\hat{\rho},\hat{H}_{m}]-\gamma /2(\hat{C}^\dagger
\hat{C} \hat{\rho}+ \hat{\rho} \hat{C}^\dagger \hat{C}-
2\hat{C}^\dagger \hat{\rho} \hat{C} ),
\end{equation} with $ \hat{C}=\hat{a}_\uparrow \hat{a}_\downarrow $ the operator that
destroys a pair atoms in different spin states and which account for
inelastic spin relaxation processes  across the resonance. Here the
constant $\gamma$ is related to the two body loss rate density,
$\gamma \approx \frac{2K_2}{V}$, which satisfies the relation
\begin{equation}
\dot{N}=\frac{ K_2}{V} N^2.
\end{equation} $K_2$ depends on the imaginary part of $a_{\downarrow \uparrow}$
as $16 \pi {\rm Im}(a_{\downarrow \uparrow})/ m$ and  ${\rm
im}(a_{\downarrow \uparrow})$ varies with the  applied magnetic
field, the  difference in magnetic moment between  the Feshbach
resonance state  and a pair of atoms in the entrance channel,
$\mu_{res}$ and the  inverse molecular lifetime,
$\gamma_{res}=1/\tau_{res}$, as \cite{Julienne}:

\begin{equation}
 {\rm im}(a_{\downarrow \uparrow})=a_{\downarrow
\uparrow}^{bg} \Delta B \frac{ \gamma_{res}/(4
\mu_{res})}{(B-B_o)^2+[\gamma_{res}/(4 \mu_{res})]^2}.
\end{equation}

In Fig.5 we summarize our numerical solutions by  plotting the
scaling exponent $b$ of the optimal sensitivity with the initial
number of atoms, $\log [\delta \chi|_{opt}/\chi] =- b \log[N(t=0)] +
{\rm const}$, versus $\gamma/\chi$ for both initially uncorrelated
atoms and initially squeezed atoms. The plot demonstrates the
drastic reduction of the sensitivity with $\gamma$. For both
initially uncorrelated atoms and initially squeezed atoms one
requires to limit $\gamma/\chi<0.03$ to have at least shot noise
resolution, i.e $b=1/2$.

Experimentally both two and three body particle losses contribute.
As three body losses scale as the four power of the scattering
length \cite{Cornell} they decrease faster than two-body losses away
from the resonance. As such, the sensitivity of the method can be
optimized by,  on one hand, tuning the magnetic field far enough
from the resonance so that mainly two-body losses are relevant  and,
on the other, by reducing $K_2$ by   using a very narrow Feshbach
resonance, and by considering species of atoms that belong to the
lowest magnetic manifold so they  posses long relaxation times.

\begin{figure}[tbh]
\begin{center}
\leavevmode {\includegraphics[width=2.5 in]{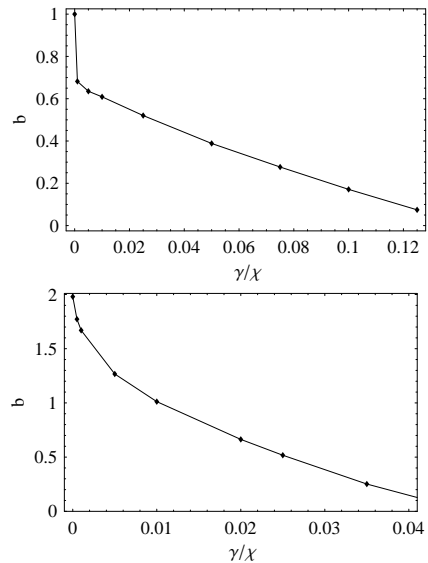}}
\end{center}
\caption{ Scaling exponent $b$ of the optimal sensitivity  with the
initial number of atoms, $\log [\delta \chi|_{opt}/\chi] =- b
\log[N(t=0)] + {\rm const}$, vs. $\gamma/\chi$ for both initially
uncorrelated atoms (top) and initially squeezed atoms (bottom).}
\end{figure}

\subsection{Fermionic atoms in optical lattices}

Recently the dynamics of   bosonic atoms with two relevant internal
states loaded  in a deep optical lattice has been used to perform
precision measurements of atomic scattering properties
\cite{Widera1}. Explicitly, the experiment detected  the
modification in the Ramsey fringes caused by  frequency shifts
induced by interatomic interactions in  wells occupied with two
atoms, and used it to measure the elastic scattering length
difference. As the various wells in the optical lattice behave as
independent wells due to the suppressed tunneling between them, the
experimental phase sensitivity  was limited to the shot noise
resolution associated  to statistical fluctuations in a finite
number of wells. In practice however, the sensitivity  was much
lower due to the two-body losses close to the Feshbach resonance in
the double occupied wells.

We now discuss how to generalize these ideas by using the collective
spin dynamics (instead of the single well one) of fermionic atoms in
an optical lattice loaded with one atom per site. Specifically we
demonstrate how to engineer an effective $J_z^2$ Hamiltonian in such
systems. The latter  can be  used to perform measurements of the
scattering properties via  the entanglement dynamics but    without
the limitation of particle losses at the resonance.

The effective   dynamics of a system of spinor ultra-cold atoms
confined in a unit filled
 lattice, deep in the Mott Insulator regime \cite{Fisher}
can be described in terms of effective spin operators,
$\hat{\sigma}_{j}^{z}=\hat{n}_{\uparrow j}- \hat{n}_{\downarrow j}$,
$\hat{\sigma}_{j}^{+}=\hat{\sigma}_{j}^{x}+
i\hat{\sigma}_{j}^{y}=\hat{a}_{\uparrow
j}^{\dagger}\hat{a}_{\downarrow j}$, where $\hat{a}_{\sigma,j}$ are
the annihilation operator of a particle of type
$\sigma=\uparrow,\downarrow$  at site $j$,  and $\hat{n}_{\sigma
j}=\hat{a}_{\sigma,j}^\dagger \hat{a}_{\sigma,j}$ are number
operators. In such spin variables the effective Hamiltonian maps to
a spin XXZ Hamiltonian \cite{Duan} :

\begin{eqnarray}
\hat{H}_{lat}&=&\hat{H}_{H}+\hat{H}_{I},\\
\hat{H}_{H}&=&\sum_{\left\langle i,j\right\rangle } \bar{\lambda}
\vec{\sigma} _{i}\cdot \vec{\sigma}_{j},\\
\hat{H}_{I}&=&\sum_{\left\langle i,j\right\rangle }\bar{\chi}\sigma
_{i}^{z}\sigma _{j}^{z}, \end{eqnarray} Here $\left\langle
i,j\right\rangle$ means that the sum is over nearest neighbors,
$t_{\sigma }$ are spin-dependent tunneling energies and $U_{\sigma
\sigma' }$  are on-site interaction energies proportional to the
 s-wave scattering lengths between the various components. For
fermions $U_{\uparrow \uparrow },U_{\downarrow \downarrow }\gg
U_{\uparrow \downarrow }$ due to the Pauli exclusion principle. The
Heisenberg, $\hat{H}_H$, and Ising $\hat{H}_I$, coupling constants
are given by

\begin{eqnarray}
\bar{\lambda} =\pm \frac{ t_{\uparrow }t_{\downarrow }}{U_{\uparrow
\downarrow }}, \quad\quad \bar{\chi}=\frac{\left( t_{\uparrow }\mp
t_{\downarrow }\right) ^{2}}{2U_{\uparrow \downarrow }}-\frac{
t_{\uparrow } ^{2}}{U_{\uparrow \uparrow }}-\frac{ t_{\downarrow }
^{2}}{U_{\downarrow  \downarrow }}, \end{eqnarray} where the upper
and lower signs are for fermionic and bosonic atoms respectively.

Using spin dependent optical potentials one can tune the various
coupling constants and in particular one can engineer the condition
$\bar{\lambda} \gg \bar{\chi }$. In this limit, we may treat
  the effect of the Ising term, $\hat{H}_I$, by means of perturbation theory.
 Assuming that at $t=0$ the initial state is prepared
within the $J=N/2$ manifold, $\mathcal{P}$,  a  perturbative
analysis predicts
 that for times $t$ such that $\bar{\chi}t<\bar{\lambda}/\bar{\chi}$,
 $\hat{H}_{H}$ confines the dynamics to $\mathcal{P}$  and
transitions outside it   can be neglected. As a consequence, only
the projection of  $\hat{H}_{I}$  on it, which corresponds   to
\begin{equation}
{\rm Tr}_\mathcal{P} \hat{H}_I= {\chi }\hat{J}_{z}^{2}+ {\rm const}
\quad \quad {\chi }\equiv \frac{2 z \bar{\chi}}{N-1} \label{impl}
\end{equation} is effective  and $H_I$ acts as a long range
Hamiltonian \cite{gap}. In Eq. (\ref{impl})  $z$ is the number of
nearest neighbors. With this effective Hamiltonian the spectroscopic
method described above can be applied in lattice systems to measure
interactions.

Moreover, due to the fact that the large Heisenberg term  restricts
the dynamics to the $\mathcal{P}$ manifold, $\hat{H}_H$ acts as an
effective $ \hat{J}^{2}$. Therefore, by adding a $\Omega_0
\hat{J}_x$ term via a Raman transition with effective two-photon
Rabi frequency $\Omega_0$ \cite{zoller} one can also implement in
such lattice systems  Eq. (\ref{Hadiabatic}). This Hamiltonian
allows, as described above, for the squeezed state preparation.

The realization  of $ \hat{J}_z^{2}$ in unit filled optical lattices
has  the strong advantage that two and three body losses are
suppressed and thus they are not longer  a limitation in these
systems. It is important to note, however, that only fermionic atoms
can be used if $\chi$ is going to be tuned by means of a Feshbach
resonance. The reason is that for bosons, as
$a_{\uparrow\downarrow}$ grows, $\chi$ also grows and at some point
the condition $\bar{\lambda} \gg \bar{\chi }$ is violated. On the
other hand, for fermions, as $a_{\uparrow\downarrow}$ grows, the
ratio $\bar{\lambda}/ \bar{\chi }$ remains constant.

The drawback of the  lattice implementation  is the factor of $N$ in
the denominator of the effective $\chi$, because the slower dynamics
limits the sensitivity of our method.  Typical experiments that use
lattices deep enough for the effective spin model to be valid, work
in a parameter regime where
 $t_\uparrow
t_\downarrow/U_{\uparrow\downarrow} \sim  10^{-1}{\rm kHz}-1{\rm
kHz}$. Using spin dependent optical lattices \cite{zoller} one can
control the  tunneling rates and set $(t_\uparrow- t_\downarrow)^2
\sim 0.1 t_\uparrow t_\downarrow$. Additionally  close to the
resonance $U_{\uparrow\downarrow}$ can be enhanced up to about  $10$
times its off resonance value \cite{Cornell}. Therefore the revival
time in these set-ups varies with the magnetic field between
$N\times$ ($10$ ms - $10^2$ms). Thus, in order to keep the
measurement time in a reasonable experimental time scale, one should
limit the experiment to 1D lattice systems which have of the order
of $20$ atoms per tube \cite{Paredes}. The scalability problem
certainly limits the  achievable sensitivity. Nevertheless, even
with these reduced number of atoms, if initially squeezed atoms are
used, the phase accuracy $\delta\chi/\chi \propto \delta
a_{\uparrow\downarrow}/a_{\uparrow\downarrow}\sim 10^{-2}$ is within
the  $1\%$ sensitivity  required to  test  the proton-electron mass
ratio variation on the level of $10^{-11}$,
  in a narrow Feshbach resonance ($\sim 1$ mG)\cite{Flambaum}.

\section{Conclusions}

In summary, here we proposed an inteferometric method  that relays
on the quantum dynamics of interacting spins  to perform  precision
measurements of the scattering  properties with a resolution
fundamentally limited to $N^{-2}$. We discussed the class of
squeezed states that are required to achieve Heisenberg sensitivity
and proposed  a method for its generation.  We studied  the effect
of decoherence and showed that  it affects the achievable
sensitivity even with initially uncorrelated atoms as they become
vulnerable due the many-particle entanglement builded during the
many-body dynamics. We also showed that the dynamically induced
quantum  correlations help to keep the sensitivity of initially
squeezed states above the uncorrelated atom threshold even with
decoherence. This has to be contrasted with standard Ramsey
spectroscopy with non-interacting atoms where all the potential gain
in sensitivity due to initial quantum correlation is lost in the
presence of decoherence.

Finally we discussed possible physical implementation of the
proposed spectroscopy using spinor condensates and fermions in unit
filled  optical lattices. We showed that particle losses close to
the Feshbach resonance limits the sensitivity in the BEC
implementation, and thus the use of  very narrow Feshbach resonances
and atomic species with a large spin relaxation time are required to
optimize the sensitivity of the method. Our implementation of the
scheme  in optical lattice set-ups is insensitive to atomic losses
but on the other is limited by the slow dynamics. Regardless of such
problems, the proposed spectroscopy is a  new scheme that takes
advantages of many-particle entanglement to perform high precision
measurements of the  interaction parameters  in spin mixtures. We
are optimistic that emerging technology in optical and magnetic
Feshbach resonances might overcome the current experimental
limitations in such systems or that new developments in the field
 might provide alternative  set-ups  for the experimental application of
the presented method and make of it an useful spectroscopic
technique.

\section{ Acknowledgements}

This work was supported by ITAMP, NSF (Career Program), AFOSR and
the David and Lucille Packard Foundation.

\end{document}